\documentclass{aa}  

\usepackage{graphicx}
\usepackage{txfonts}
\usepackage{lipsum}
\usepackage{subcaption}         
\usepackage{lscape}             
\usepackage{placeins}           
\usepackage{xcolor}  
\usepackage{soul}

\begin{document}

\title{Deep Learning-Based Coarse Alignment and Cophasing of a Distributed-Aperture Telescope}
\subtitle{Application to the Small ExoLife Finder (SELF)}

%
%
%

   \author{N. Arteaga-Marrero\inst{1,2}\corrauth{narteaga@iac.es}        
        \and J. Iborra-Luis\inst{1,2}\email{jordi.iborra@iac.es}
        \and A. Padrón-Brito\inst{1,2}\email{maria.auxiliadora.padron@iac.es}
        \and J. Kuhn\inst{1,2,3,4}\email{jeff.reykuhn@yahoo.com}
        }

   \institute{Instituto de Astrofísica de Canarias, Laboratory for Innovation in Opto-Mechanics (LIOM), Calle Vía Láctea s/n, E-38200 La Laguna, Tenerife, Spain
   \and Universidad de la Laguna (ULL), Departamento de Astrofísica, E-38206 La Laguna, Tenerife, Spain
   \and Institute for Astronomy, University of Hawai’i, 2680 Woodlawn Dr. Honolulu, HI, 96822, USA
   \and MorphOptic, Inc. , 2540 Kekaulike Ave, Kula, HI, 96790, USA}

   \date{Received September 30, 20XX}

 
  \abstract
   {
The ExoLife Finder (ELF), a 35--50-meter-class Fizeau interferometric telescope, was designed to detect potential biosignatures on exoplanets through direct imaging. The Small ExoLife Finder (SELF), a 3.5-meter prototype, serves as a testbed for the innovative technologies required to realize such a facility.}
   {
A key challenge in SELF is the precise cophasing of its distributed aperture, where the optical path differences between subapertures must remain below a small fraction of the operating wavelength under dynamic conditions. This work investigates the use of convolutional neural networks (CNNs) to estimate alignment errors directly from focal-plane images, providing the basis for fast, data-driven coarse-alignment routine capable of reducing misalignments to the few-micrometer level.}
   {
A supervised regression framework was developed using a custom CNN trained to model the nonlinear mapping between subaperture misalignments and focal-plane intensity distributions. Training and validation datasets were generated from high-fidelity optical simulations of the simplified SELF system. Gaussian-noise augmentation was used to assess robustness under realistic observing conditions. Several established CNN architectures were also evaluated for comparison.}
   {
The proposed approach successfully reconstructed piston and tilt misalignments across all mirror pairs, demonstrating that focal-plane interference patterns can be effectively exploited for alignment estimation. The custom CNN provided a favorable balance between estimation performance, robustness, and computational efficiency, with inference times on the order of milliseconds, compared with several standard CNN architectures. A trade-off between estimation performance and noise robustness was identified, with Gaussian-noise augmentation improving reconstruction at low signal-to-noise ratios while reducing it under near noise-free conditions. }
   { 
These results demonstrate the feasibility of data-driven, focal-plane-based wavefront sensing and alignment control for distributed-aperture telescopes. The proposed framework, validated on a high-fidelity optical model, enables autonomous and computationally efficient cophasing strategies for future high-resolution interferometric systems.}


   \maketitle

\section{Introduction}
The ExoLife Finder (ELF) is a 35--50-meter-class hybrid interferometric telescope concept aimed at the direct imaging of exoplanets and the detection of potential biosignatures \citep{kuhn2025creating}. Its feasibility relies on several key technological developments currently under investigation at the Laboratory for Innovation in Opto-Mechanics (LIOM) at the Instituto de Astrofísica de Canarias. These developments are being validated using a 3.5-meter prototype telescope, hereafter referred to as the Small ExoLife Finder (SELF) \citep{kuhn2025creating, kuhn2022small}.

SELF features a hybrid architecture that combines conventional imaging with Fizeau interferometry. Its distributed aperture consists of 15 primary and 15 secondary mirrors arranged in a one-to-one configuration. The beams from each primary--secondary pair must be coherently combined at the focal plane to form a diffraction-limited image. Achieving and maintaining precise cophasing is therefore crucial, particularly under dynamic conditions induced by gravity loading, thermal perturbations, atmospheric turbulence, and mechanical vibration. In practice, the optical path difference (OPD) between aperture pairs must be controlled to within a small fraction of the operating wavelength. At a wavelength of 1~$\mu$m, this corresponds to stabilization at the level of a few tens of nanometers.

The distributed design of SELF introduces significant challenges for cophasing its subapertures. In large segmented telescopes and sparse-aperture imaging systems, cophasing is traditionally achieved using dedicated wavefront sensing techniques. These are commonly implemented in the pupil plane, such as Shack-Hartmann \citep{platt2001history} or pyramid sensors \citep{ragazzoni1996pupil}, or directly in the focal plane through methods such as phase diversity \citep{paxman1992joint} and interferometric fringe analysis \citep{chanan1998phasing}. More recently, machine learning (ML) approaches have also been explored to infer wavefront information from focal-plane images. These include phase diversity-based methods \citep{orban2021focal}, coronagraph-enabled sensing \citep{quesnel2022deep}, and deep learning methods capable of estimating cophasing errors from a single image \citep{cheng2026deep, ma_piston_2019, dumont2024phasing}. 

High-precision cophasing of a distributed-aperture system using a ResNet18 architecture trained on a single far-field image was previously reported \citep{cheng2026deep}. A single alignment degree of freedom (DOF) was considered under fixed noise conditions, demonstrating the feasibility of recovering piston information with root mean square error (RMSE) values of 0.517~nm in simulation and 2.9~nm experimentally. More generally, data-driven alignment methods based on neural networks have become increasingly popular for learning the complex mapping between image features and alignment parameters, as they enable near real-time feedback with millisecond-scale inference latency without requiring external sensors \citep{nie2026deep}. 

Focal-plane approaches are particularly attractive as they operate on the science image itself, reducing non-common path aberrations (NCPAs), which are a key limitation in high-contrast imaging \citep{martinez_speckle_2012, guyon2018extreme, savransky2012focalplane}. However, these methods tipically assume small wavefront errors and therefore rely on a prior coarse cophasing stage. In segmented telescopes, this is typically achieved using edge sensing systems, whereas in sparse-aperture architectures such as SELF, coarse cophasing must rely on optical methods. 

Furthermore, most existing studies rely on simplified Fourier-optics models or analytical point spread function (PSF) representations \citep{orban2021focal, ma_piston_2019}, which do not fully capture the complexity of real optical systems. They also typically consider only a limited number of alignment DOFs, most commonly piston, rather than the coupled tilt and piston misalignments encountered in realistic distributed-aperture telescopes. Consequently, it remains unclear how well existing CNN architectures generalize to realistic distributed-aperture systems involving multiple coupled alignment DOFs.

SELF presents a non-redundant distributed aperture, where cophasing information can, in principle, be directly inferred from focal-plane images under ideal conditions without complex non-linear algorithms \citep{baron2008unambiguous}. This has motivated the use of focal-plane wavefront sensing \citep{kuhn2025creating, kuhn2022small}, optionally complemented by photonic-based sensors \citep{padron2024co, padron2026focal}, in both cases combined with ML approaches. While these methods have demonstrated promising results, key challenges remain during the design and validation stages, particularly in identifying the relevant DOF for each primary--secondary pair and understanding the coupling between perturbations.

To address these challenges, a supervised ML regression framework was developed based on a high-fidelity optical model implemented in Ansys Zemax OpticStudio \citep{zemax2024}. A custom convolutional neural network (CNN) was trained to learn the nonlinear mapping between subaperture misalignments and their corresponding focal-plane signatures, thereby establishing a coarse-alignment capability to the few-micrometer level, within the capture range required for subsequent nanometer-level cophasing. To the authors’ knowledge, this represents the first study to apply deep learning to a high-fidelity ray-tracing model of a distributed-aperture telescope for multi-DOF alignment estimation. 

To reduce the complexity of the full system while preserving the essential multi-aperture interference structure, a simplified configuration consisting of four primary--secondary pairs was adopted, with variations restricted to the primary-mirror DOF. This configuration is consistent with the initial experimental tests planned for SELF, enabling future laboratory validation of the proposed approach.

The paper is organized as follows. Section 2 introduces the methodology, including the high-fidelity optical model, the data generation procedure, a brief theoretical motivation, and the CNN-based regression framework. Section 3 presents the results and analyzes the impact of data augmentation and noise robustness, as well as compares the proposed framework with established CNN architectures. Section 4 discusses the results and outlines future research directions. Finally, Section 5 summarizes the main conclusions.

\section{Methodology}
\subsection{Zemax Telescope Model and Datasets}
\label{sec:ZemaxModel}
A high-fidelity optical model of a four-aperture configuration of SELF was implemented in Ansys Zemax OpticStudio \citep{zemax2024}. The model consisted of four primary--secondary mirror pairs, each forming sections of a classical Gregorian telescope. The primary mirrors had a diameter of 0.5~m and were defined to be identical off-axis segments of a common parent paraboloid. Each primary was paired with a 20~mm diameter off-axis ellipsoidal secondary mirror, from a common parent ellipsoid, forming an image at the Gregorian focus. The system was designed with a narrow field of view of 10~arcseconds, an effective focal length of 32.5~m, and a focal ratio of F/9.3 \citep{kuhn2025creating, kuhn2022small}.

The primary mirrors were arranged in a circular array with a diameter of 3.5~m, consistent with the SELF design. The four apertures were located at asymmetric azimuthal angles of $0^{\circ}$, $72^{\circ}$, $192^{\circ}$, and $264^{\circ}$ relative to the array center, with the corresponding primary--secondary combinations hereafter referred to as pairs 1--4. This asymmetric configuration was chosen to reduce phase-retrieval degeneracies by providing distinct baseline orientations between pairs. The computed focal-plane images were sampled on a $512 \times 512$ pixel grid, corresponding to a $1024 \times 1024$~$\mu$m physical array.

\begin{figure}[ht!]
\centering
\includegraphics[width=\hsize]{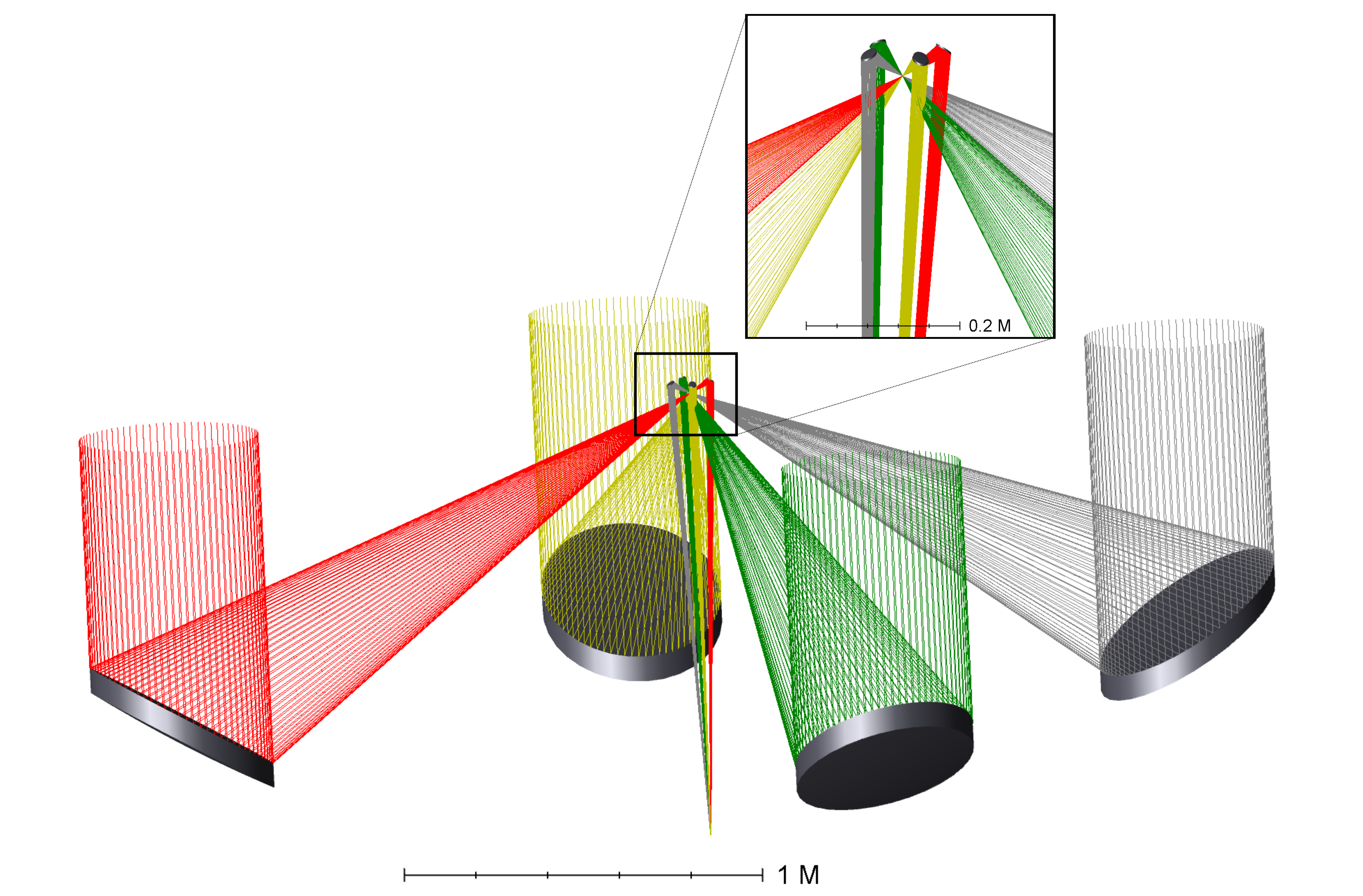}
\caption{Three-dimensional layout of the four-aperture SELF optical model implemented in Ansys Zemax OpticStudio \citep{zemax2024}, showing the arrangement of the primary and secondary mirrors, together with the optical path used for focal-plane image generation. The inset provides a magnified view of the secondary mirrors.}
\label{fig:Zemax_3DLayout}
\end{figure}

Figure~\ref{fig:Zemax_3DLayout} illustrates the implemented model, including the configuration of the four primary--secondary pairs and the optical path used for focal-plane image generation. 

The focal-plane response of the system was computed using the Huygens PSF algorithm in Ansys Zemax OpticStudio, which combines ray tracing with wave-optics propagation. Controlled perturbations were applied to the primary-mirror DOFs, and the corresponding point spread functions (PSFs) were computed. In this approach, a grid of rays was launched through the optical system, with each ray representing a wavelet characterized by a specific amplitude and phase. The diffraction intensity at any point in the image plane was obtained as the squared magnitude of the coherent sum of these wavelets. This procedure was repeated across the image-plane grid to construct the full PSF. 

\begin{figure}[ht!]
\centering
\includegraphics[width=0.5\hsize]{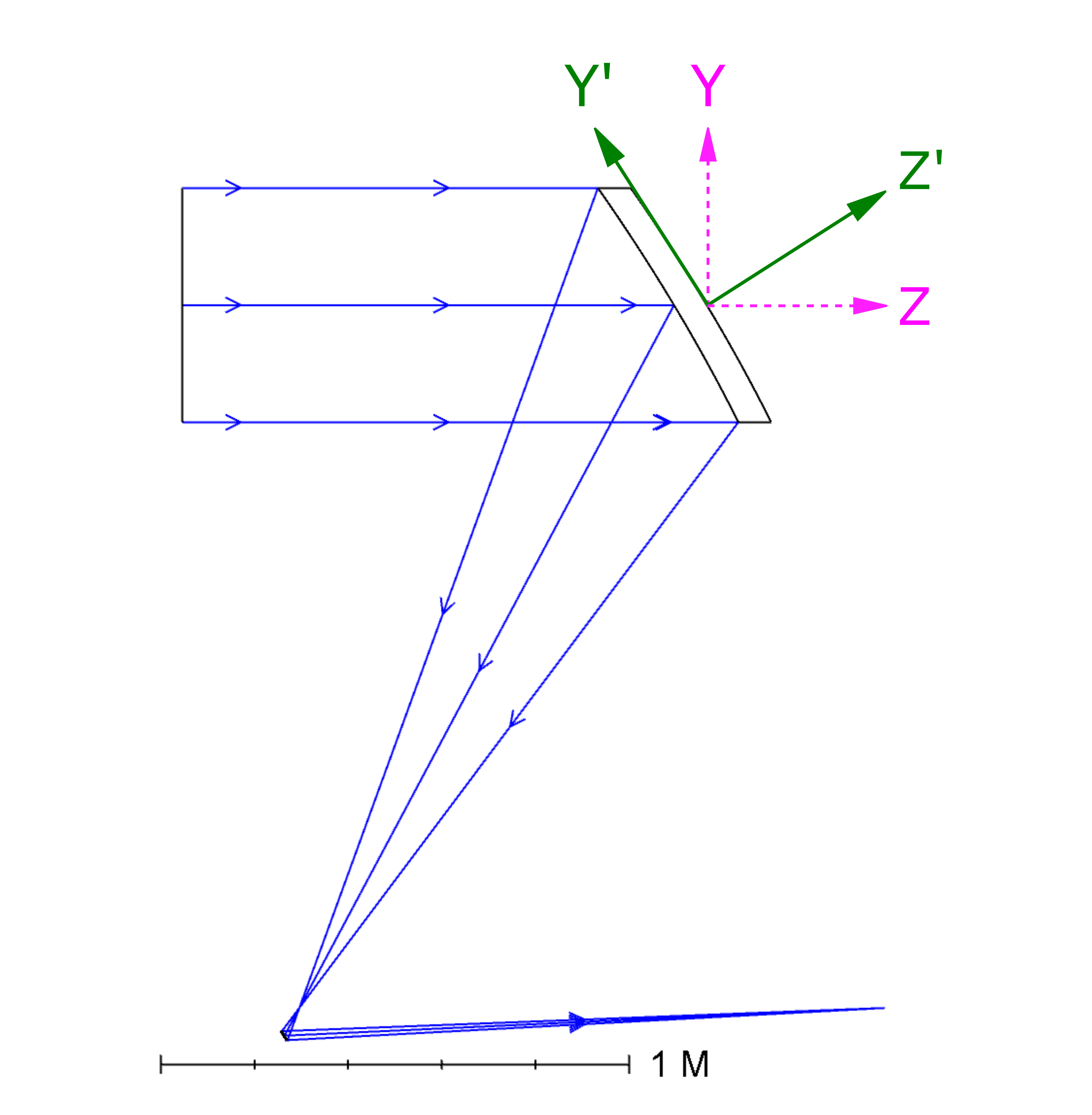}
\caption{Local coordinate system used to define perturbations applied to each primary mirror (green arrows), with the Z-axis aligned with the mirror surface normal. The global coordinate system is shown for reference (pink arrows). }
\label{fig:Zemax_coordinates}
\end{figure}

The perturbations were defined in a local reference frame centered on each primary mirror, with the Z-axis aligned with the surface normal, as shown in Figure~\ref{fig:Zemax_coordinates}. In this proof-of-concept study, the DOFs considered for each primary mirror were piston and tilts about the X and Y axes. Accordingly, piston corresponded to a translation along the Z-axis, i.e., along the surface normal.

Zemax was accessed via its Python API, enabling automated dataset generation. A custom Python script was therefore developed to interface with the optical model. Random perturbations were applied to the DOFs of each primary mirror before computing the corresponding focal-plane image. The perturbation ranges were $\pm 50~\mu$m for piston and $\pm 10~\mu$rad for tilt. These ranges were selected to represent realistic alignment perturbations expected during operation. The selected operating wavelength was set to 1~$\mu$m.

A dataset of 10000 samples was generated by simultaneously applying random perturbations to all primary mirrors. For independent evaluation, a separate test set comprising 1000 samples was generated using the same procedure. This dataset was subsequently modified by adding Gaussian white noise to the PSFs generated in Ansys Zemax OpticStudio, enabling evaluation under varying noise conditions.

\subsection{Image Formation in a Distributed-Aperture Telescope}
The focal-plane intensity distribution of a coherent imaging system is determined by the generalized pupil function. Assuming monochromatic coherent illumination and the Fraunhofer approximation, the generalized pupil function is defined as 

\begin{equation}
\Psi(\mathbf{r})
=
P(\mathbf{r})
\exp\!\left[i\phi(\mathbf{r})\right],
\label{eq:generalizedPupil}
\end{equation}

\noindent where $P(\mathbf{r})$ denotes the pupil transmission function and $\phi(\mathbf{r})$ represents the wavefront phase. The corresponding PSF is then given by the squared magnitude of the Fourier transform of the generalized pupil function \citep{goodman1969introduction, baron2008unambiguous},

\begin{equation}
I(\boldsymbol{\theta})
=
\left|
\mathcal{F}
\left\{
\Psi(\mathbf{r})
\right\}
\right|^{2},
\label{eq:psf}
\end{equation}

\noindent where $\mathcal{F}\{\cdot\}$ denotes the Fourier transform and $\boldsymbol{\theta}$ represents the angular coordinates in the image plane.

For a distributed-aperture telescope, the generalized pupil function can be expressed as the coherent sum of the complex fields associated with the individual subapertures,


\begin{equation}
\Psi(\mathbf{r})
=
\sum_{k=1}^{N}
P_k(\mathbf{r}-\mathbf{r}_k)
\exp\!\left[i\phi_k(\mathbf{r}-\mathbf{r}_k)\right],
\label{eq:DistributedAperture}
\end{equation}

\noindent where $P_k(\mathbf{r}-\mathbf{r}_k)$ denotes the transmission function of the $k$-th subaperture centered at position $\mathbf{r}_k$, and $\phi_k(\mathbf{r}-\mathbf{r}_k)$ is the corresponding subaperture wavefront phase. Consequently, piston and tilt perturbations modify the phase distribution across the generalized pupil, producing measurable changes in the interference pattern at the focal plane \citep{baron2008unambiguous}.

The phase associated with each subaperture depends on its alignment state. Restricting the analysis to piston and tilt perturbations, the phase distribution over the $k$-th subaperture can be approximated by


\begin{equation}
\phi_k(x-x_k,y-y_k)
=
\frac{2\pi}{\lambda}
\left(
p_k
+t_{x,k}(x-x_k)
+t_{y,k}(y-y_k)
\right),
\label{eq:Phase}
\end{equation}

\noindent where $p_k$ denotes the piston displacement, $t_{x,k}$ and $t_{y,k}$ are the wavefront tilt coefficients along the local $x$- and $y$-directions, respectively, and $\lambda$ is the operating wavelength.

The focal-plane intensity is a nonlinear function of the piston and tilt perturbations across the distributed aperture. Consequently, estimating the subaperture alignment errors from focal-plane images constitutes an inherently nonlinear inverse problem, motivating the use of data-driven regression models such as CNNs. The preceding equations provide the physical basis for the image formation process. However, the datasets used in this work were generated using high-fidelity ray-tracing simulations in Ansys Zemax OpticStudio \citep{zemax2024}, which account for the complete optical geometry of the SELF system, and therefore, extend beyond the idealized Fourier optics formulation.

\subsection{CNN-Based Regression Framework}
\label{sec:ML_Model}
The ML approach was formulated as a regression problem aimed at estimating the magnitudes and signs of the perturbations applied to each primary aperture. The model took as input the focal-plane intensity distribution, represented by PSFs. The outputs, serving as labels in the supervised learning framework, corresponded to the ground-truth piston and tilt perturbations applied to each primary mirror. These estimates can be used to derive the corrective commands required to restore the system to its nominal alignment. 

The implemented regression model employed a compact architecture designed to extract relevant features from the focal-plane images in the generated dataset. The neural network, hereafter referred to as the custom CNN, consisted of four two-dimensional convolutional layers, each followed by instance normalization, Rectified Linear Unit (ReLU) activations, max pooling, and 25\% dropout. These layers were followed by two fully connected layers. The first fully connected layer included layer normalization, a ReLU activation, and 25\% dropout, while the second predicted the perturbations applied. 

The network predicted 12 output parameters corresponding to piston and tilt perturbations for each of the four primary mirrors: four piston values and eight tilt values (X and Y tilts for each mirror).

The dataset was split into training and test sets (80/20\%). The training set was then used in a 10-fold cross-validation procedure, in which 90\% of the samples were used for training and 10\% for validation at each fold. A fixed random seed was used to ensure reproducibility. The 20\% test split was kept noise-free and used for clean evaluation, while the separate 1000-sample held-out dataset was used exclusively for noise robustness evaluation.

Mean squared error (MSE) was used as the loss function, and the network was optimized using the Adam optimizer \citep{kingma2014adam} with a learning rate of 0.001 and a batch size of 16. Early stopping, with a patience of 10 epochs, was employed to prevent overfitting, together with model checkpointing to save the weights whenever the validation loss improved. For each fold, the model corresponding to the lowest validation loss was retained and evaluated on the independent test set. 

Performance metrics obtained on the test set were aggregated across the 10 folds and reported as the mean and standard deviation for each measure, including the RMSE, mean absolute error (MAE), and coefficient of determination ($R^2$).

Training and inference were performed on the IACTEC computational system, equipped with an Intel\textsuperscript{\textregistered} Core\texttrademark{} Ultra 9 185H CPU (2.5~GHz), 32~GB of system RAM, and an NVIDIA Quadro RTX 4000 GPU with 8~GB of dedicated memory. 

\subsection{Data Augmentation}
Deep learning approaches generally require large training datasets. However, PSF generation based on high-fidelity optical simulations is computationally intensive \citep{nie2026deep}. For the simulated optical system considered here, dataset generation required approximately 1.2 minutes per sample. Accordingly, data augmentation was employed to increase the effective size of the training dataset and to assess its impact on model performance.

The augmentation procedure consisted of adding Gaussian white noise to the PSFs generated in Ansys Zemax OpticStudio \citep{zemax2024}, thereby generating additional noisy realizations of the original PSFs to simulate camera readout noise and increase the number of training instances without requiring additional optical simulations. Signal-to-noise ratios (SNRs) ranging from 20 to 40~dB were considered. This augmentation was applied to different fractions of the training dataset (0.25, 0.5, 0.75, and 1.0), corresponding to progressively larger numbers of augmented training instances, with a separate CNN model trained for each configuration.

The augmentation process was stochastic and applied on the fly during training, ensuring variability across epochs. Consequently, computational cost increased with the fraction of augmented data, as reflected in the training time.

\subsection{Performance Comparison with Established Architectures}
The custom CNN was compared with established architectures such as MobileNetV3 \citep{howard2019searching}, EfficientNetB0 \citep{tan2019efficientnet}, ResNet18 \citep{he2016deep}, and VGG16 \citep{simonyan2014very}. All architectures were trained from scratch, without pretrained weights to ensure a fair comparison.

The input layers of these architectures were modified, where necessary, to accept grayscale images. In addition, the original prediction stage of each architecture was adapted for the regression task by replacing it with two fully connected layers, mirroring those used in the custom CNN. Specifically, the first fully connected layer included layer normalization, a ReLU activation, and 25\% dropout, while the second predicted the 12 output parameters corresponding to the piston and tilt perturbations.

The dataset was split and the models were trained following the same procedure described for the custom CNN, including the 80/20 train–test split and 10-fold cross-validation. The loss function, optimizer, batch size, and learning rate were kept unchanged, except for the VGG16 architecture, for which the learning rate was reduced to 0.0001 to ensure stable convergence. The same performance metrics were used for evaluation.

The selected models were also evaluated for robustness using the independent 1000-sample held-out dataset, to which Gaussian white noise with SNRs ranging from 10 to 50~dB was added.

\section{Results}
\subsection{Impact of Data Augmentation}
The performance of the regression model as a function of the augmentation fraction is summarized in Tables~\ref{Zemax_tilt_aug} and \ref{Zemax_piston_aug} for tilt and piston estimation, respectively. Here, the augmentation fraction denotes the proportion of training samples randomly augmented with Gaussian white noise during training. Training and average inference times are reported in Table~\ref{Zemax_tilt_aug} and apply to both tables. As expected, the training time increased with the augmentation fraction, from 2.4~h without augmentation to 18.9~h when all training samples were eligible for augmentation.

\begin{figure}[ht!]
\centering
\includegraphics[width=\hsize]{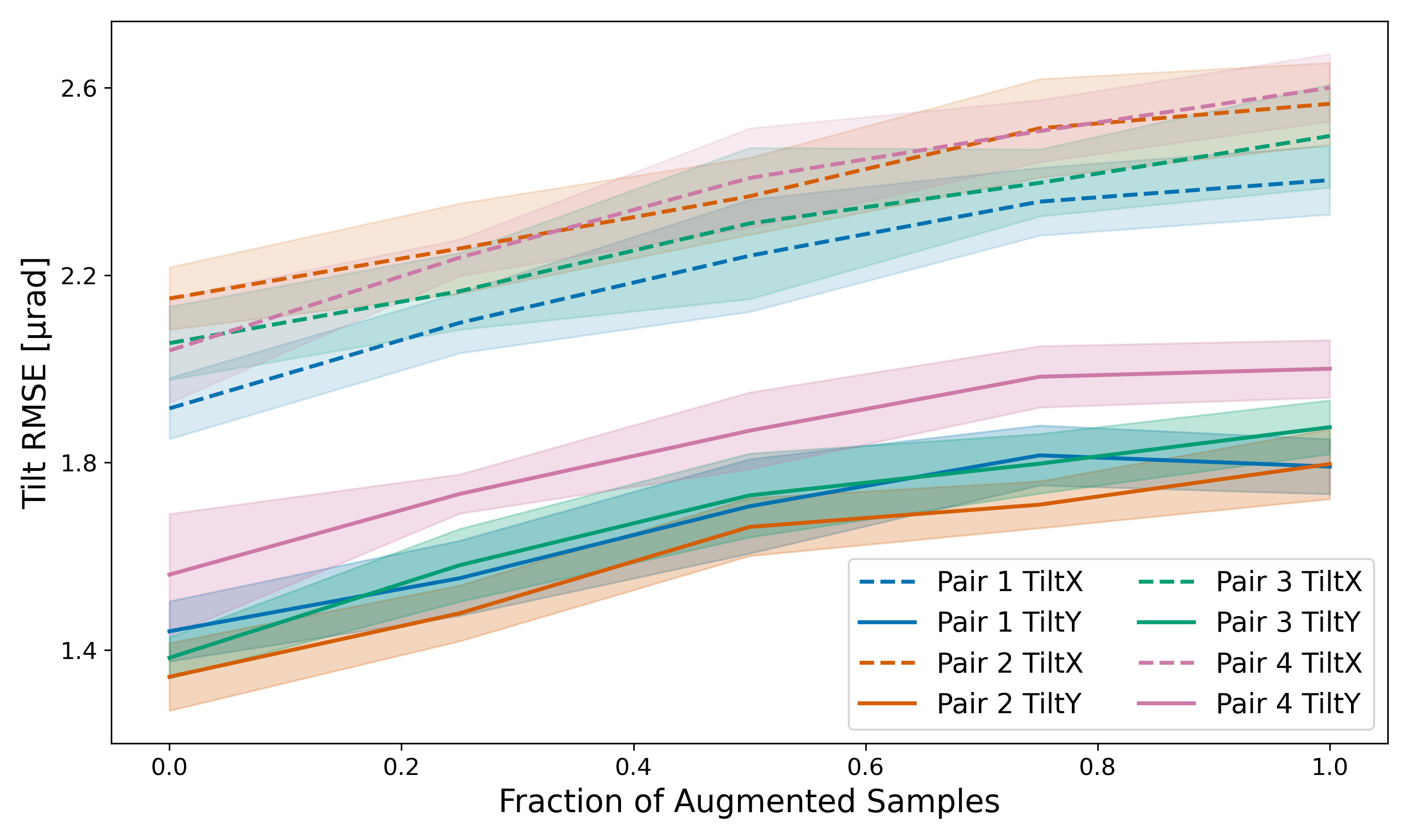}
\caption{Root mean square error (RMSE) of the tilt estimates along the X (dashed) and Y (solid) axes as a function of the augmentation fraction for all mirror pairs. Lines indicate the mean over the 10 cross-validation folds, and the shaded regions represent one standard deviation.}
\label{fig:tilt_augmentation}
\end{figure}

\begin{figure}[ht!]
\centering
\includegraphics[width=\hsize]{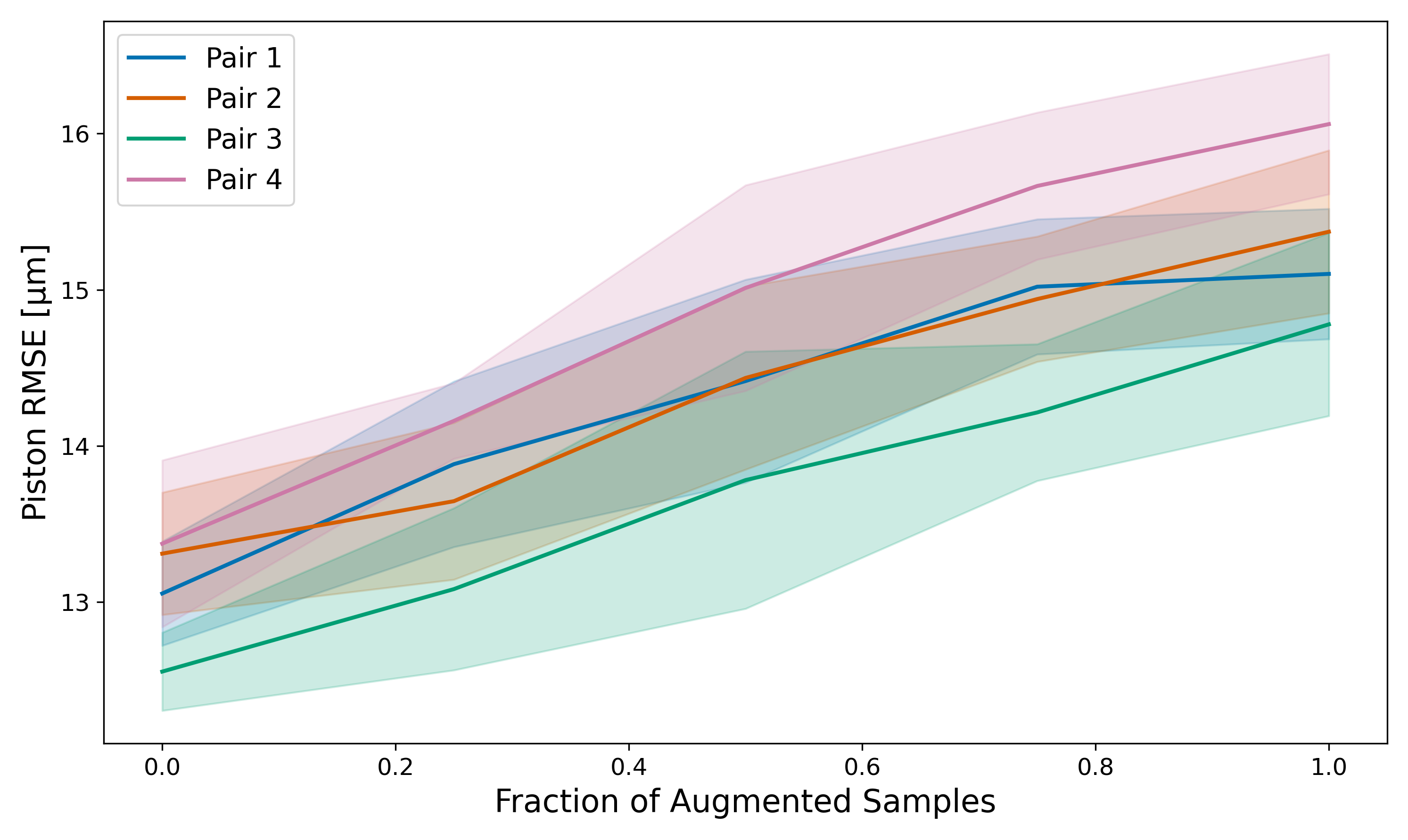}
\caption{Root mean square error (RMSE) of the piston estimates for each mirror pair as a function of the augmentation fraction. Lines indicate the mean over the 10 cross-validation folds, and the shaded regions represent one standard deviation.}
\label{fig:pistondiff_augmentation}
\end{figure}

Figure~\ref{fig:tilt_augmentation} shows the variation of the tilt RMSE values as a function of the augmentation fraction, while Figure~\ref{fig:pistondiff_augmentation} presents the corresponding results for piston. 

Increasing the augmentation fraction resulted in a systematic increase in the estimation errors for both piston and tilt. Since the test set remained noise-free at this stage, this degradation most likely reflected the mismatch between the noisy training distribution and the clean evaluation data. As expected, training with increasingly noisy inputs reduced the model's ability to optimize performance under ideal observing conditions.

Furthermore, tangential tilt estimation (X axis) consistently yielded larger errors than the sagittal tilt estimation (Y axis). This asymmetry likely reflected differences in the information content of the corresponding PSFs, which appeared to encode more discriminative features along the Y axis. For piston, reconstruction errors also differed consistently between mirror pairs, indicating that the optical response was not identical for all aperture configurations.

Statistical analysis using the Friedman test \citep{friedman1937} confirmed significant differences ($p < 0.05$) in prediction performance between mirror pairs for all DOFs and augmentation levels, indicating that the observed variations were not random. Post-hoc Wilcoxon tests \citep{wilcoxon1945individual} with Holm correction \citep{holm1979simple} further showed that the degree of separation depended on both the estimated variable and the augmentation fraction. 

Overall, piston exhibited the most stable and discriminative behavior across mirror pairs, whereas tilt along the Y axis showed moderate separability, and tilt along the X axis was the most sensitive DOF to the augmentation strategy. Moderate augmentation generally improved the statistical separability of the mirror responses, particularly for piston, although this did not translate into improved estimation performance on the clean test set. 

\longtab[1]{
\footnotesize
\setlength{\tabcolsep}{3pt}
\begin{longtable}{llllllll}
\caption{Regression performance for tilt estimation as a function of the augmentation fraction, defined as the proportion of training samples randomly augmented with Gaussian white noise during training. Results are reported as the mean ± standard deviation over the 10 cross-validation folds. Performance is quantified using the root mean square error (RMSE), mean absolute error (MAE), and coefficient of determination ($R^2$). Training and average inference times are also reported.}\\
\label{Zemax_tilt_aug}\\
\hline\hline
Augmentation Fraction & Training/Validation & Train (h) & Infer (ms) & Tilt & RMSE ($\mu$rad) & MAE ($\mu$rad) & $R^2$ \\
\hline
\endhead
    0               & 7200 / 800    & 2.4   & 0.128 & Y (Pair 1) & 1.440 ± 0.064 & 1.009 ± 0.048 & 0.939 ± 0.006   \\
                    &               &       &       & X (Pair 1) & 1.916 ± 0.065 & 1.457 ± 0.052 & 0.889 ± 0.008   \\

                    &               &       &       & Y (Pair 2) & 1.342 ± 0.072 & 0.929 ± 0.056  & 0.945 ± 0.006   \\
                    &               &       &       & X (Pair 2) & 2.150 ± 0.067 & 1.656 ± 0.063  & 0.860 ± 0.009  \\
 
                    &               &       &       & Y (Pair 3) & 1.383 ± 0.044 & 0.958 ± 0.037  & 0.942 ± 0.004   \\
                    &               &       &       & X (Pair 3) & 2.055 ± 0.078 & 1.572 ± 0.066  & 0.873 ± 0.010   \\
 
                    &               &       &       & Y (Pair 4) & 1.561 ± 0.130 & 1.072 ± 0.089  & 0.926 ± 0.013   \\
                    &               &       &       & X (Pair 4) & 2.039 ± 0.112 & 1.524 ± 0.089  & 0.873 ± 0.014   \\

\hline
0.25                & 9000 / 1000   & 8.4   & 0.140 & Y (Pair 1) & 1.553 ± 0.080 & 1.053 ± 0.066 & 0.929 ± 0.007   \\
&      &       &       & X (Pair 1) & 2.098 ± 0.065 & 1.583 ± 0.043 & 0.867 ± 0.008   \\

&    &       &       & Y (Pair 2) & 1.478 ± 0.060 & 1.018 ± 0.046 & 0.934 ± 0.005  \\
&                    &       &       & X (Pair 2) & 2.257 ± 0.096 & 1.734 ± 0.076 & 0.845 ± 0.013   \\

&                    &       &       & Y (Pair 3) & 1.581 ± 0.078 & 1.086 ± 0.065 & 0.924 ± 0.007   \\
&                    &       &       & X (Pair 3) & 2.166 ± 0.082 & 1.635 ± 0.065 & 0.859 ± 0.011   \\

&                    &       &       & Y (Pair 4) & 1.733 ± 0.042 & 1.151 ± 0.027 & 0.909 ± 0.004   \\
&                    &       &       & X (Pair 4) & 2.237 ± 0.039 & 1.670 ± 0.038 & 0.847 ± 0.005   \\

\hline
0.50                & 10800 / 1200  & 13.0  & 0.210 & Y (Pair 1) & 1.707 ± 0.101 & 1.151 ± 0.066  & 0.914 ± 0.010   \\
                    &               &       &       & X (Pair 1) & 2.241 ± 0.120 & 1.688 ± 0.089  & 0.848 ± 0.017   \\

                    &               &       &       & Y (Pair 2) & 1.662 ± 0.062 & 1.121 ± 0.060  & 0.916 ± 0.006   \\
                    &               &       &       & X (Pair 2) & 2.368 ± 0.082 & 1.818 ± 0.069  & 0.830 ± 0.012   \\
          
                    &               &       &       & Y (Pair 3) & 1.730 ± 0.089 & 1.153 ± 0.065  & 0.909 ± 0.010   \\
                    &               &       &       & X (Pair 3) & 2.310 ± 0.162 & 1.747 ± 0.133  & 0.839 ± 0.024   \\
         
                    &               &       &       & Y (Pair 4) & 1.868 ± 0.082 & 1.231 ± 0.066  & 0.894 ± 0.009   \\
                    &               &       &       & X (Pair 4) & 2.407 ± 0.106 & 1.772 ± 0.092  & 0.823 ± 0.016   \\

\hline
0.75                & 12600 / 1400  & 16.2 & 0.144  & Y (Pair 1) & 1.815 ± 0.064 & 1.226 ± 0.048 & 0.903 ± 0.007   \\
                    &               &       &       & X (Pair 1) & 2.357 ± 0.073 & 1.771 ± 0.057 & 0.832 ± 0.010   \\
        
                    &               &       &       & Y (Pair 2) & 1.710 ± 0.050 & 1.162 ± 0.030 & 0.911 ± 0.005    \\
                    &               &       &       & X (Pair 2) & 2.514 ± 0.105 & 1.926 ± 0.089 & 0.808 ± 0.016   \\
       
                    &               &       &       & Y (Pair 3) & 1.798 ± 0.064 & 1.202 ± 0.042 & 0.902 ± 0.007   \\
                    &               &       &       & X (Pair 3) & 2.397 ± 0.071 & 1.797 ± 0.045 & 0.827 ± 0.010   \\
       
                    &               &       &       & Y (Pair 4) & 1.983 ± 0.066 & 1.311 ± 0.042 & 0.881 ± 0.008   \\
                    &               &       &       & X (Pair 4) & 2.508 ± 0.066 & 1.837 ± 0.054 & 0.808 ± 0.010   \\

\hline
1.00                & 14400 / 1600   & 18.9 & 0.210  & Y (Pair 1) & 1.791 ± 0.059 & 1.205 ± 0.037 & 0.906 ± 0.006   \\
                    &                &      &        & X (Pair 1) & 2.403 ± 0.074 & 1.809 ± 0.062 & 0.825 ± 0.010    \\

                    &                &       &       & Y (Pair 2) & 1.797 ± 0.075 & 1.203 ± 0.049 & 0.902 ± 0.008   \\
                    &                &       &       & X (Pair 2) & 2.566 ± 0.088 & 1.982 ± 0.072 & 0.800 ± 0.014   \\
     
                    &                &       &       & Y (Pair 3) & 1.875 ± 0.058 & 1.240 ± 0.047 & 0.893 ± 0.007   \\
                    &                &       &       & X (Pair 3) & 2.497 ± 0.110 & 1.893 ± 0.095 & 0.812 ± 0.017   \\               
         
                    &                &       &       & Y (Pair 4) & 2.000 ± 0.061 & 1.324 ± 0.044 & 0.879 ± 0.007   \\
                    &                &       &       & X (Pair 4) & 2.600 ± 0.072 & 1.921 ± 0.061 & 0.794 ± 0.011   \\

\hline
\end{longtable}
}

\longtab[1]{
\begin{longtable}{llllll}
\caption{Regression performance for piston estimation as a function of the augmentation fraction. Results are reported as the mean ± standard deviation over the 10 cross-validation folds. Performance is quantified using the root mean square error (RMSE), mean absolute error (MAE), and coefficient of determination ($R^2$). Training and inference times are reported in Table~\ref{Zemax_tilt_aug}.}\\
\label{Zemax_piston_aug}\\
\hline\hline
Augmentation Fraction & Training/Validation & Piston & RMSE ($\mu$m) & MAE ($\mu$m) & $R^2$ \\
\hline
\endhead
    0               & 7200 / 800    & Pair 1 & 13.055 ± 0.334 & 10.022 ± 0.303 & 0.798 ± 0.011 \\
                    &               & Pair 2 & 13.310 ± 0.392 & 10.260 ± 0.316 & 0.785 ± 0.013 \\
                    &               & Pair 3 & 12.555 ± 0.250 & 9.648  ± 0.214 & 0.812 ± 0.008  \\
                    &               & Pair 4 & 13.374 ± 0.534 & 10.219 ± 0.421 & 0.779 ± 0.018 \\
                    
\hline
 0.25               & 9000 / 1000   & Pair 1 & 13.883 ± 0.530 & 10.654 ± 0.450 & 0.772 ± 0.018 \\
                    &               & Pair 2 & 13.646 ± 0.502 & 10.504 ± 0.411 & 0.774 ± 0.017 \\
                    &               & Pair 3 & 13.083 ± 0.519 & 9.974  ± 0.447 & 0.795 ± 0.016 \\
                    &               & Pair 4 & 14.160 ± 0.241 & 10.820 ± 0.256 & 0.752 ± 0.008  \\

\hline
 0.50               & 10800 / 1200  & Pair 1 & 14.415 ± 0.650 & 11.085 ± 0.548 & 0.754 ± 0.023 \\
                    &               & Pair 2 & 14.435 ± 0.586 & 11.160 ± 0.460 & 0.747 ± 0.021 \\
                    &               & Pair 3 & 13.781 ± 0.823 & 10.547 ± 0.691 & 0.772 ± 0.028 \\
                    &               & Pair 4 & 15.011 ± 0.659 & 11.467 ± 0.536 & 0.721 ± 0.024 \\

\hline
 0.75               & 12600 / 1400  & Pair 1 & 15.019 ± 0.433 & 11.483 ± 0.356 & 0.733 ± 0.015 \\
                    &               & Pair 2 & 14.940 ± 0.401 & 11.548 ± 0.351 & 0.729 ± 0.015 \\
                    &               & Pair 3 & 14.214 ± 0.438 & 10.852 ± 0.285 & 0.758 ± 0.015 \\
                    &               & Pair 4 & 15.664 ± 0.471 & 11.944 ± 0.426 & 0.697 ± 0.018 \\

\hline
 1.00               & 14400 / 1600  & Pair 1 & 15.101 ± 0.417 & 11.635 ± 0.383 & 0.730 ± 0.015 \\
                    &               & Pair 2 & 15.371 ± 0.522 & 11.966 ± 0.376 & 0.713 ± 0.020 \\
                    &               & Pair 3 & 14.778 ± 0.586 & 11.337 ± 0.471 & 0.739 ± 0.021 \\
                    &               & Pair 4 & 16.060 ± 0.449 & 12.297 ± 0.403 & 0.681 ± 0.018 \\
\hline
\end{longtable}
}

\subsection{Noise Robustness}
For the noise robustness analysis, the model achieving the best validation performance, for each augmentation fraction, across the 10 folds was selected and evaluated on the independent 1000-sample held-out dataset. Gaussian white noise was added independently to all 1000 samples at each of the considered SNR levels. These levels included both the interval used during training (20--40~dB) and additional values between 10 and 50~dB to assess performance within and beyond the training regime.

\begin{figure}[ht!]
\centering
\includegraphics[width=\hsize]{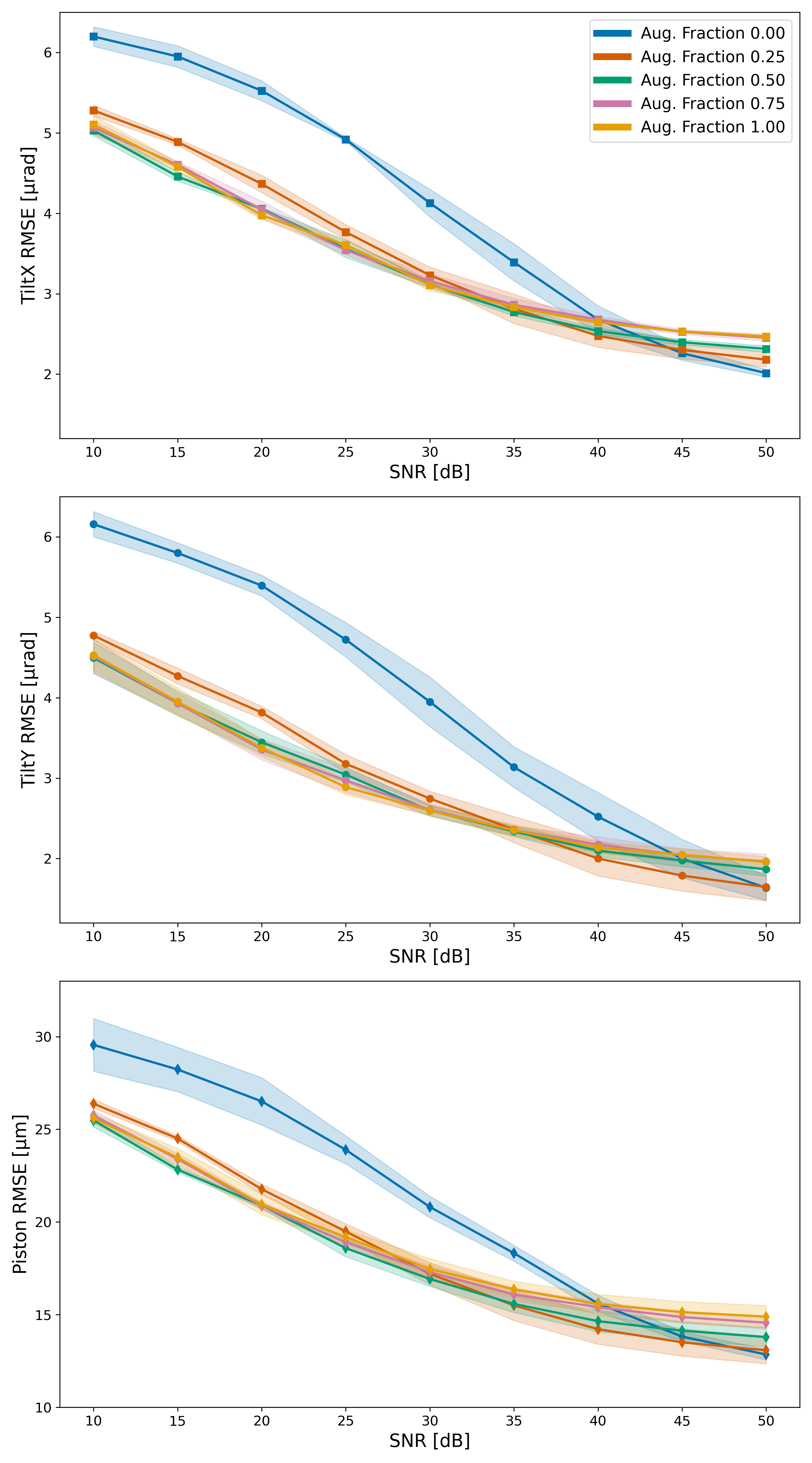}
\caption{Average root mean square error (RMSE) for tilt estimation along the X (top) and Y (middle) axes, and piston estimation (bottom), as a function of the signal-to-noise ratio (SNR) for models trained with different augmentation fractions. Lines represent the mean RMSE, and shaded regions indicate one standard deviation across the four mirror pairs. For each mirror pair, the RMSE was computed independently over the held-out test dataset.}
\label{fig:RMSE_noise_all}
\end{figure}

The results are summarized in Figure~\ref{fig:RMSE_noise_all}, which shows the average RMSE as a function of SNR for each trained model. As expected, performance improved with increasing SNR across all models, reflecting the reduced impact of noise on the input data. 

Overall, a clear trade-off was observed between robustness and performance. Models trained with higher augmentation fractions exhibited lower errors at low SNRs, indicating improved robustness to noisy inputs. In contrast, the model trained without augmentation achieved the lowest errors under near noise-free conditions. These results indicate that data augmentation improved robustness to input noise at the expense of reduced performance on clean data.

Within the SNR range used during training (20--40~dB), the optimal augmentation strategy depended on the noise level. At 20~dB, models trained with higher augmentation fractions achieved the lowest errors (full augmentation for tilt and 0.5 augmentation for piston). At 40~dB, lower augmentation fractions (approximately 0.25) consistently produced the best performance across all DOF. The same trend persisted outside the training regime. At SNRs below 20~dB, the fully augmented model achieved the lowest prediction errors, whereas above 40~dB the model trained without augmentation consistently performed best.

In summary, data augmentation did not improve performance on clean data, but substantially increased robustness under noisy conditions, revealing an inherent trade-off between nominal performance and noise tolerance. These results suggest that the effectiveness of data augmentation depends on the expected observing conditions and highlight the importance of matching the training data distribution to the operational environment.

\subsection{Comparison with Established CNN Architectures}
The custom CNN architecture was intentionally kept shallow to prioritize inference speed, thereby providing a practical balance between computational efficiency and reconstruction performance. For comparison purposes, several established CNN architectures were also evaluated. Table~\ref{CNN_comparison} summarizes their performance in terms of trainable parameters, training time, inference latency, and reconstruction accuracy. No data augmentation was applied in these experiments.

\begin{figure}[ht!]
\centering
\includegraphics[width=\hsize]{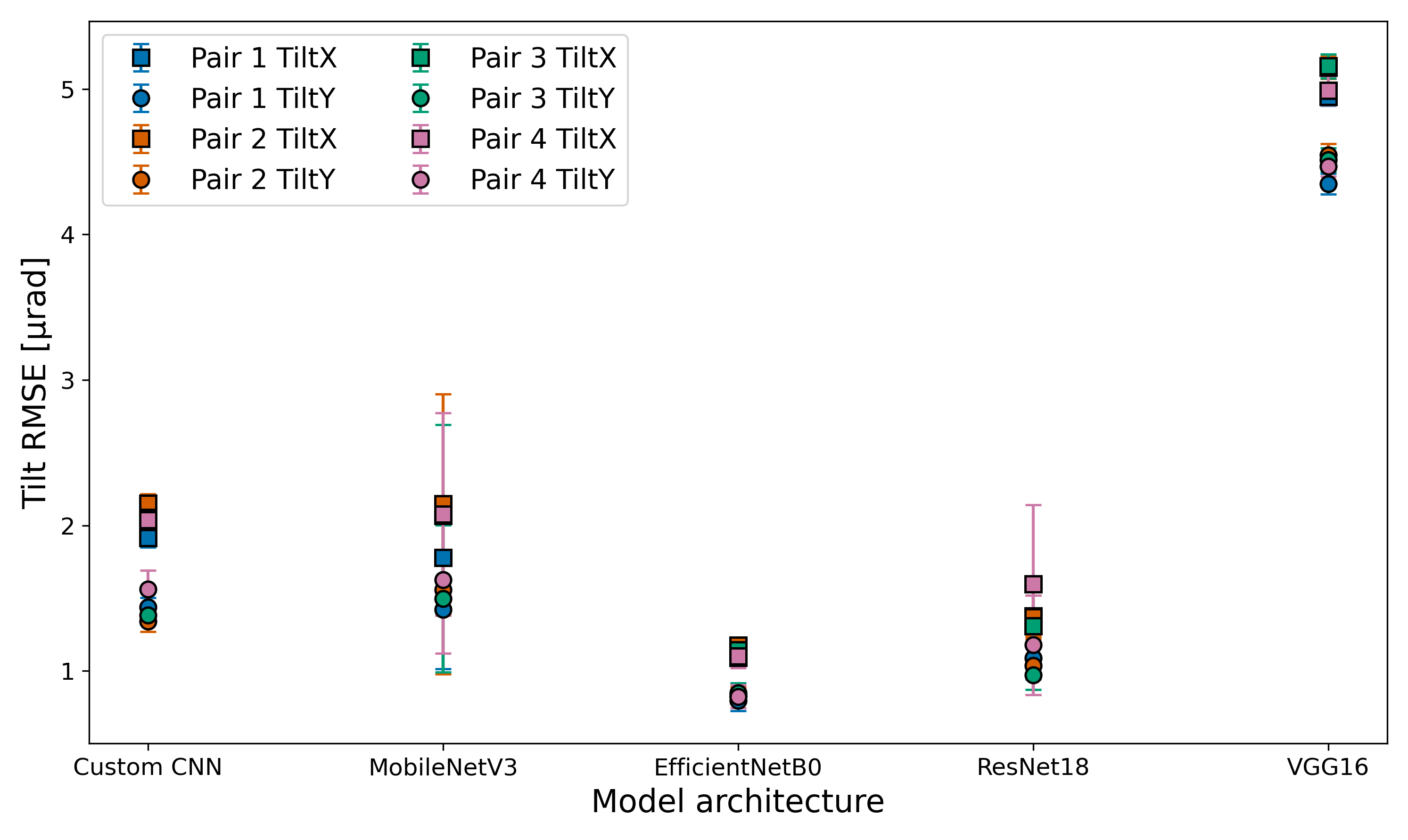}
\includegraphics[width=\hsize]{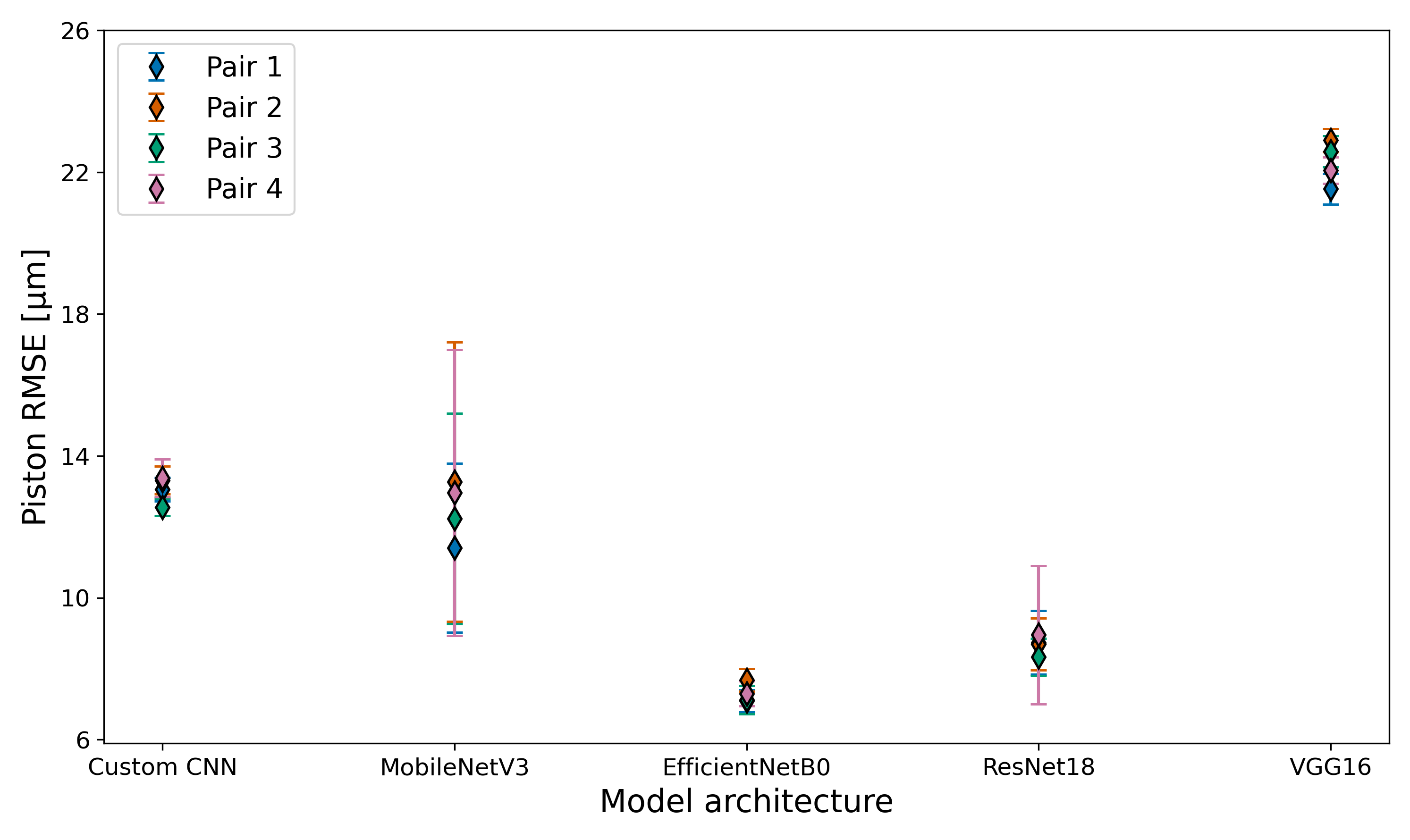}
\caption{Comparison of regression performance across the evaluated CNN architectures. The top panel shows the RMSE for tilt estimation along the X and Y axes, while the bottom panel shows the RMSE for piston estimation. Error bars represent one standard deviation across the 10 cross-validation folds.}
\label{fig:RMSE_CNN_comp_all}
\end{figure}

Figure~\ref{fig:RMSE_CNN_comp_all} shows the performance of each architecture in the absence of data augmentation, providing a baseline comparison. Overall, the results revealed a trade-off between reconstruction performance and computational efficiency. 

EfficientNetB0 achieved the lowest RMSE values for both tilt and piston estimation. However, this gain came at the cost of increased inference latency, which may limit its suitability in time-constrained scenarios. The custom CNN achieved slightly higher errors while maintaining one of the lowest inference latencies, making it an attractive compromise for real-time applications. VGG16 exhibited comparable variability but substantially larger reconstruction errors.

The statistical analyses indicated that the inferred reconstruction errors depended on both the estimated DOF and the selected CNN architecture. Although the custom CNN and VGG16 generally exhibited more consistent pairwise differences between mirror configurations than the remaining architectures, these patterns varied across the predicted DOF and did not reveal a common physical trend. Consequently, the principal comparison focused on the overall reconstruction and computational performance of the different architectures.

\begin{figure}[ht!]
\centering
\includegraphics[width=\hsize]{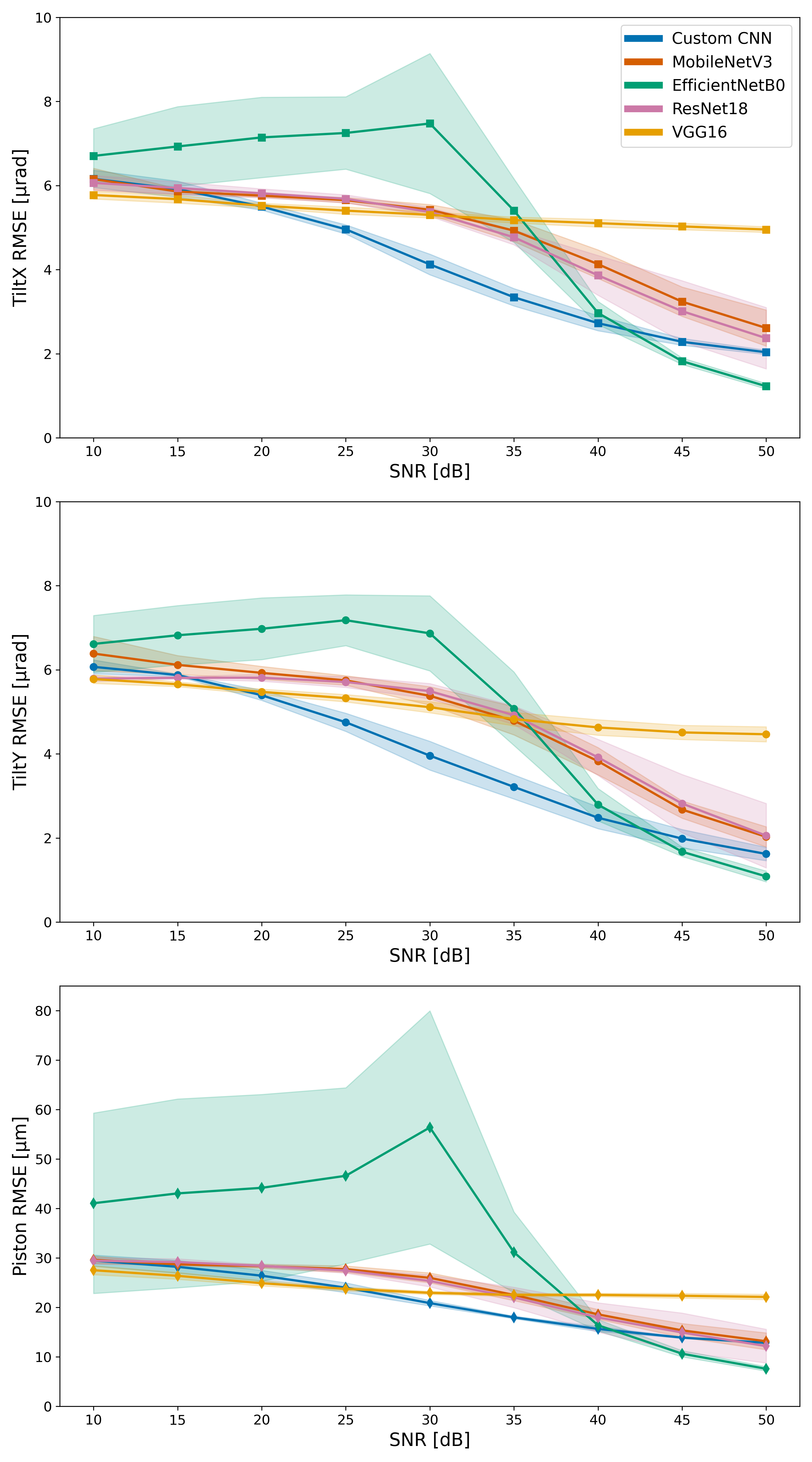}
\caption{Regression performance of the evaluated CNN architectures as a function of the signal-to-noise ratio (SNR). The top and middle panels show the RMSE for tilt estimation along the X and Y axes, respectively, while the bottom panel shows the RMSE for piston estimation. Lines represent the mean RMSE and shaded regions indicate one standard deviation across the four mirror pairs. For each mirror pair, the RMSE was computed independently over the held-out test dataset.}

\label{fig:RMSE_CNN_comp_noise}
\end{figure}

Figure~\ref{fig:RMSE_CNN_comp_noise} compares the robustness of the evaluated architectures under varying noise levels. Within the SNR range used during training (20--40~dB), the custom CNN provided the lowest errors for tilt estimation, whereas performed similarly to VGG16 for piston estimation. Below the training range (SNR < 20~dB), VGG16 consistently achieved the best performance across all estimated DOF. At high SNR values (>40~dB), EfficientNetB0 produced the lowest reconstruction errors, although its performance degraded more rapidly as the noise level increased.

\longtab[1]{
\begin{longtable}{lllllll}
\caption{Comparison of the evaluated CNN architectures in terms of trainable parameters, training time, inference latency, and reconstruction accuracy. Reconstruction performance is reported as the mean $\pm$ standard deviation of the RMSE across mirror pairs for each DOF.}\\
\label{CNN_comparison}\\
\hline\hline
Architecture & Params (M) & Train (h) & Infer (ms) & TiltX RMSE ($\mu$rad) & TiltY RMSE ($\mu$rad) & Piston RMSE ($\mu$m) \\
\hline
\endhead
Custom CNN      & 1.2   & 2.1   & 0.128 & 2.040 ± 0.084 & 1.431 ± 0.082 & 13.074 ± 0.322 \\
MobileNetV3     & 1.1   & 1.4   & 0.906 & 2.018 ± 0.141 & 1.527 ± 0.075 & 12.465 ± 0.719 \\
EfficientNetB0  & 4.3   & 1.8   & 0.828 & 1.128 ± 0.034 & 0.826 ± 0.019 & 7.297  ± 0.236 \\
ResNet18        & 11.4  & 0.9   & 0.311 & 1.414 ± 0.110 & 1.069 ± 0.076 & 8.675  ± 0.224 \\
VGG16           & 16.8  & 1.0   & 0.132 & 5.060 ± 0.094 & 4.469 ± 0.076 & 22.263 ± 0.524 \\
\hline
\end{longtable}
}

In summary, although EfficientNetB0 achieved the lowest reconstruction errors under high-SNR conditions, the custom CNN provided the best performance for multi-DOF estimation within the SNR range used during training (20--40~dB). Additionally, this architecture had a substantially lower number of trainable parameters and exhibited one of the lowest inference latencies among the evaluated architectures. Thus, the custom CNN represents a practical choice for real-time wavefront estimation.

 \section{Discussion}
The proposed CNN-based regression framework demonstrated the feasibility of estimating multiple alignment DOFs directly from focal-plane images using a high-fidelity optical model of SELF. The achieved inference times indicate that the approach is compatible with future near real-time active alignment systems. In the context of the SELF alignment strategy, the proposed approach is intended to address the coarse-alignment stage, rapidly reducing large initial alignment errors to the few-micrometer level and bringing the system within the capture range of the subsequent precise speckle-based cophasing process required for nanometer-level optical coherence. The physically realistic simulations also enabled the development and validation of alignment strategies prior to laboratory implementation.

The use of a ray-tracing model increased the computational cost of dataset generation but produced physically realistic PSFs that accurately represented the optical geometry of the telescope. Gaussian white noise was added to a fraction of the focal-plane images during training to better approximate realistic observing conditions. The resulting dataset enabled the custom CNN to successfully estimate both tilt and piston perturbations directly from focal-plane images.

Across the evaluated dataset configurations, the proposed approach achieved low estimation errors for both tilt and piston perturbations. The method remained robust over a range of perturbation magnitudes, although its sensitivity to measurement noise depended on the adopted data augmentation strategy. Increasing the augmentation fraction improved robustness at low SNR by reducing estimation errors but degraded performance under nearly noise-free conditions, highlighting a trade-off between noise robustness and performance on clean data. Consequently, the optimal augmentation level depended on the expected operating conditions.

Statistical analysis further showed that the DOFs responded differently to noise and data augmentation. In particular, piston exhibited the most stable and interpretable PSF variations across mirror pairs.

It is important to note that the objective of the present study was not final nanometre-level cophasing, but rather the estimation of large initial alignment errors during the coarse alignment stage. The reported piston values therefore correspond to rigid-body translations of the primary mirrors along their local surface normals, rather than a pure optical path difference as in classical interferometric cophasing. Consequently, the standard limitation of global piston indetermination from focal-plane images did not directly apply in this context. These perturbations were therefore substantially larger than the precision required for coherent operation. Fine cophasing would subsequently primarily concern the piston difference between mirror pairs rather than the global piston. Multi-wavelength measurements could provide additional information for estimating the relative piston between mirrors and may therefore help achieve the nanometre-level precision required for coherent operation.

The systematically higher estimation errors observed for piston compared with tilt suggest that the corresponding image features were more difficult for the network to exploit. Unlike tilt perturbations, translations of the primary mirrors along their local surface normals produced more subtle changes in the focal-plane intensity distribution, resulting in a less distinctive mapping between the images and the underlying alignment parameters. Although these image features may arise from changes in the relative optical path between apertures, they may also reflect secondary effects introduced by the optical geometry of the system. A more detailed analysis of the physical information exploited by the network is therefore required to determine the dominant mechanisms underlying piston estimation.

Independently of the architecture used, tilt reconstruction along the Y axis was consistently more accurate than along the X axis, indicating that the information encoded in the PSFs exhibited anisotropic discriminative features. This asymmetry may reflect intrinsic properties of the optical system, whereby the interference patterns encoded more structured information along the Y direction. Although the overall estimation error depended on the selected CNN architecture, this directional asymmetry remained consistent, suggesting that it originated from the optical encoding of the PSFs rather than from the network design. Nevertheless, differences between architectures indicated that the inferred distinguishability of individual mirror pairs could depend on the representational capacity of the model.

The relative performance of the evaluated architectures depended on the operating SNR. The custom CNN and VGG16 exhibited the greatest robustness under noisy conditions, whereas EfficientNetB0 achieved the lowest estimation errors under nearly noise-free conditions. When inference latency, estimation performance, and robustness were considered jointly, the custom CNN provided the most balanced compromise for the present application, making it a suitable candidate for future near real-time active alignment systems. Its favorable performance suggests that a lightweight architecture with sufficient representational capacity to capture the mid-scale spatial variations of the interference patterns was adequate for this task. With approximately 1.2 million trainable parameters, compared with 4.3–16.8~M for the reference architectures, the custom CNN achieved competitive performance while maintaining low computational cost.

The present results extend previous demonstrations of CNN-based piston estimation from focal-plane images \citep{cheng2026deep} to the simultaneous estimation of multiple alignment DOFs during the coarse alignment stage using a high-fidelity optical model. Direct comparison of the reported RMSE values is not meaningful because of the different estimation objectives, optical configurations, and perturbation ranges. Nevertheless, the results suggest that the suitability of a given CNN architecture depends on both the complexity of the optical system and the specific alignment task being addressed.

Overall, this study demonstrated that CNNs trained on physically realistic ray-tracing simulations can estimate multiple alignment DOFs directly from focal-plane images. The proposed framework provides a practical foundation for data-driven coarse alignment strategies for distributed-aperture telescopes such as SELF, with the potential to rapidly reduce initial alignment errors to the few-micrometer level and bring the system within the capture range of subsequent precise speckle-based cophasing for coherent system operation.

\section{Conclusions}
This study demonstrated a CNN-based framework for the alignment control of a distributed-aperture telescope, showing that CNNs can effectively learn the nonlinear mapping between focal-plane intensity distributions and subaperture misalignment parameters. The framework was validated using a high-fidelity optical model of a simplified four-aperture configuration of SELF, achieving tilt reconstruction errors below 2.6~$\mu$rad RMSE and piston reconstruction errors below 16~$\mu$m RMSE across the evaluated mirror pairs and augmentation conditions.

The results showed that data augmentation played a key role in balancing reconstruction performance and robustness to noise across different observational regimes. The comparison of CNN architectures further demonstrated that network design influenced both estimation performance and computational efficiency. In particular, lightweight, task-specific architectures, such as the custom CNN, emerged as an efficient alternative to deeper standard models under real-time constraints, providing a favorable balance between reconstruction performance, noise robustness, and sub-millisecond inference latency, making it a promising approach for future active alignment applications. Overall, these findings emphasised that both system complexity and network architecture must be jointly considered when developing deep-learning-based alignment control strategies.

Future work will focus on experimental validation using the SELF telescope, followed by on-sky demonstrations. The framework will also be extended to include broadband illumination as well as additional alignment DOFs, including secondary mirror perturbations, together with the development of closed-loop alignment correction strategies. These efforts will advance the proposed approach toward deployment in large-scale interferometric systems such as ELF.

\begin{acknowledgements}
The authors acknowledge funding from the European Union under the ERA Chair grant (Project Ref.: 101087032). LIOM project's R\&D\&i activities are also supported by the Cabildo Insular de Tenerife thanks to the ``Apoyo a las actuaciones I+D+I en el espacio de cooperación IACTEC" collaboration agreement. The views and opinions expressed are however those of the authors only and do not necessarily reflect those of the EU or the EU Research Executive Agency. Neither the European Union nor the granting authority can be held responsible for them. 
This work is part of grant CEX2025-001609-S, awarded to the Instituto de Astrofísica de Canarias under the Severo Ochoa Centre of Excellence program and funded by MICIU/AEI/10.13039/501100011033.
The authors are grateful to the team at the Laboratory for Innovation in Opto-Mechanics (LIOM), as well as the IAC High-Performance Computing support team and hardware facilities. The authors also thank Sagittal Optics for providing the telescope model.
\end{acknowledgements}

\bibliographystyle{bibtex/aa}
\bibliography{bibtex/sample}

\end{document}